\documentclass[letter,10pt]{article}
\usepackage{hyperref}
\usepackage{color}
\usepackage{parskip}
\usepackage[affil-it]{authblk}
\usepackage{graphicx}
\usepackage[title]{appendix}
\usepackage{amssymb}
\usepackage{amsthm}
\usepackage{amsmath}
\usepackage[margin=1in]{geometry}
\usepackage{multirow}
\usepackage{subfig}

\usepackage{listings}

\definecolor{dkgreen}{rgb}{0,0.6,0}
\definecolor{gray}{rgb}{0.5,0.5,0.5}
\definecolor{mauve}{rgb}{0.58,0,0.82}
\definecolor{backcolour}{rgb}{0.97,0.97,0.97}
\title{\textsc{XtalOpt} Version 14: Variable-Composition Crystal Structure Search for Functional Materials Through Pareto Optimization}

\author[a]{Samad Hajinazar}
\author[a,*]{Eva Zurek}
\affil[a]{Department of Chemistry, State University of New York at Buffalo, Buffalo, New York  14260-3000, United States}
\affil[*]{Corresponding author: ezurek@buffalo.edu}
\date{}                     
\begin{document}

\maketitle

\begin{abstract}
Version 14 of \texttt{XtalOpt}, an evolutionary multi-objective global optimization algorithm for
crystal structure prediction, is now available for download from its official website
\href{https://xtalopt.github.io}{https://xtalopt.github.io}, and the Computer Physics Communications Library.
The new version of the code is designed to perform a ground state search for crystal structures with variable compositions by integrating a suite of \textit{ab initio} methods alongside classical and machine-learning potentials for structural relaxation.
The multi-objective search framework has been enhanced through the introduction of
Pareto optimization, enabling efficient discovery of functional materials.
Herein, we describe the newly implemented methodologies, provide detailed instructions for their use,
and present an overview of additional improvements included in the latest version of the code.
\end{abstract}



\section{Introduction}\label{sec:intro}

The development of computational methodologies and tools for the discovery of novel functional materials has taken center stage with the advent of diffusion-based models to generate crystal lattices~\cite{Alverson_2023, Luo2024, Yang_2024a, Zeni2025}, machine learning (ML) interatomic potentials to optimize these lattices and compute their energies~\cite{Zhang_2025}, as well as ML-based models for predicting material's properties~\cite{Stanev2018, 10.1021/acs.jpclett.8b00124, Goodall2020, Court2020, Hart2021, 10.1038/s41524-024-01270-1, 10.1063/5.0255385, Gan2025a}. More traditional crystal structure prediction (CSP) methods for materials discovery~\cite{Zurek:2014d, Oganov:2019a, Falls2021a, Wang2022, Conway2023} have also benefited from the myriad advances in artificial intelligence for materials. In a typical CSP search, the potential energy surface of the target chemical system has traditionally been mapped through \textit{ab initio} calculations or classical force fields, and -- more recently -- using ML potentials~\cite{SH04, SH05, SH06, Kharabadze2022, Wang2023b, Salzbrenner2023, Roberts_2024}. Then, global optimization techniques are utilized to navigate this energy landscape, while the energetics of the examined structural motifs are employed to find possible routes towards more stable phases.

Various global optimization strategies have been implemented within commonly used software platforms for CSP, such as evolutionary algorithms (\texttt{XtalOpt}~\cite{Zurek:2011a, Falls2021a}, \texttt{MAISE}~\cite{SH07}, \texttt{GASP}~\cite{GASP0}, and \texttt{USPEX}~\cite{Oganov:2006}), particle swarm optimization (\texttt{CALYPSO}~\cite{Wang:2010a}), and random structure searching (\texttt{AIRSS}~\cite{pickard2009structures}).
We have previously illustrated that the commonly used evolutionary algorithm is likely to have a higher success rate of finding the ground state structure in fixed-composition searches as compared to a purely random search \cite{Falls2021a}. However, employing constraints based upon chemical and physical intuition has made random structure searching (e.g., as implemented in the \texttt{AIRSS} code) a very powerful technique.
Further extension of the basic search algorithms to multi-objective global optimization (MOGO) approaches~\cite{Giagkiozis2015a, Gunantara2018} has facilitated the ground state search for functional materials~\cite{solomou2018, Gopakumar2018, khatamsaz2022, chen_predicting_2014, Zhang2015, NUNEZVALDEZ2018152, Maldonis2019, Cheng2020, meng_experimentally_2023, SH14}.

Another key direction in addressing the technological demand for new materials through CSP approaches involves enabling the optimization algorithm to search the composition space of a desired chemical system, i.e., a ``variable-composition" (VC) ground state search~\cite{Trimarchi:2009a, 10.1063/1.3488440, Tipton:2013b, Revard:2016a, Yang2020a, Zhu2013a, liu_prediction_2015, Kvashnin:2020a}. In particular, the advent and rapid advancements of ML interatomic potentials, including the more recent family of universal interatomic potentials (UIPs)~\cite{mace_arxiv, Deng2023a, rhodes2025orbv3, yang2024mattersim,Chen2022, Choudhary2023, GoogUniverse}, offers an enormous new opportunity for high-throughput exploration of various chemical systems in searching for (meta)stable phases of novel functional materials.

\texttt{XtalOpt}, an evolutionary algorithm designed for CSP, is a fully open-source and cross-platform software package, available in both command-line interface (CLI) and graphical user interface (GUI) modes. It is capable of performing both single-objective (SO) and multi-objective (MO) searches~\cite{SH12} for the ground state structures of crystalline systems.
Originally designed to search over a fixed chemical composition, previous implementations of \texttt{XtalOpt}~\cite{Zurek:2011a,Zurek:2011f,Zurek:2015h,Zurek:2018j,Falls2021a, SH12} start by producing an initial set of structures generated from specified supercells of the given empirical formula. Following successful local optimization of the population members, the selection of parent structures was based on the fitness, determined from the normalized relative enthalpies obtained in an SO search, or normalized relative value of user-defined objectives in a MO search~\cite{SH14,Zurek:2018j,Zurek:2019b,Wang2022a}. The offspring would be created from the parent structure(s), with the size and composition matching those of the initial user-defined list, through application of genetic operations. These genetic operations included: ``crossover" where two parent structures are mixed to generate an offspring with the same composition, as well as ``stripple" and ``permustrain" mutations that generate a new structure by applying random distortions to a selected parent structure. Filtration of the parent pool was introduced to keep only those structures with desirable characteristics~\cite{SH10}.

In the present work, we introduce the new release of \texttt{XtalOpt} (version 14) where, among various improvements, the MOGO functionality is extended by including the Pareto optimization algorithm, implementing a VC search capability that is natively integrated with the MOGO, and easy-to-use interface scripts are included for the utilization of modern UIPs in structural relaxation. These developments therefore pave the way towards an efficient ground state search for crystalline phases of materials with variable composition and desired target properties.

\begin{figure}[t]
\centering \includegraphics[width=.48\linewidth]{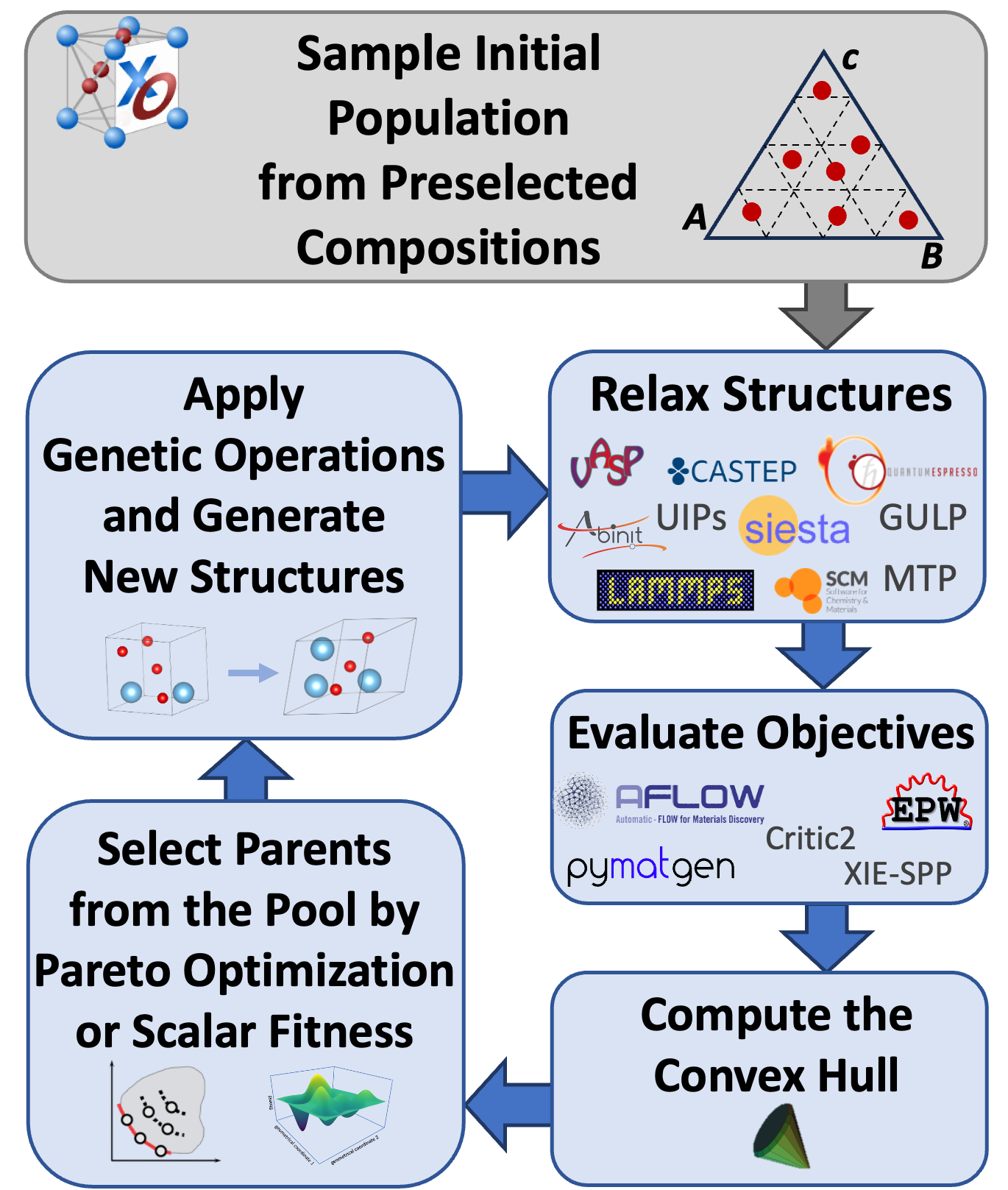}
\caption{The workflow in \texttt{XtalOpt} version 14, which offers a variable-composition multi-objective (VC-MO) search. Post relaxation, each structure's energies are used to obtain its distance above the convex hull, which is combined with the values of the user-specified objectives to determine the structure's suitability for parenting new offspring. Genetic operations are then applied to the selected parents to create new structures with compositions determined by the search type.}
\label{fig-workflow}
\end{figure}

The workflow of the \texttt{XtalOpt} VC search (optionally with multiple objectives) is illustrated schematically in Fig. \ref{fig-workflow}. The following steps are taken: (i) the initial population of structures is generated from a set of user-defined compositions and (optionally) using the \texttt{randSpg} library~\cite{Zurek:2016h}, (ii) the population members are locally optimized using one of the supported platforms, (iii) the user-specified objectives (if any) are calculated, (iv) the convex hull for the entire population is calculated and the distance above the convex hull for all successfully relaxed structures is updated via the \texttt{Qhull} library~\cite{Barber1996}, (v) the parents for new structures are selected from the pool either through Pareto optimization or by using the generalized scalar fitness function, (vi) new offspring are produced from the selected parent structures, and they are submitted for local optimization. This cycle is repeated until a pre-specified number of structures are optimized.

The implementation of the VC search functionality is made possible through a series of modifications to the \texttt{XtalOpt} algorithm, namely, using the convex hull of the population to obtain the energetic fitness of structures~\cite{Trimarchi:2009a}, expanding the existing genetic operations to perform evolutionary transformations on parent structures of (possibly) different composition and producing an offspring whose composition matches the desired search type, and adjusting the workflow for generating high-symmetry unit cells, or those that involve molecular units, etc.

The details of the implemented methodologies and instructions for using them are discussed in the following sections. While most of the described options are available in version 14.0 of \texttt{XtalOpt}, the multi-cut crossover functionality (Section \ref{sec:multi}) and the option to set a minimum total atom count (Section \ref{limit}) are introduced in versions 14.1 and 14.2, respectively.

\section{Input chemical formulas}

In the previous versions of \texttt{XtalOpt}~\cite{Zurek:2011a, SH12},
the search over a fixed chemical composition was initiated with specifying
the chemical system by its reduced empirical formula, e.g.,

{\begin{lstlisting}
empiricalFormula = Ti1O2
\end{lstlisting}}

over an optionally determined list of formula units, e.g.,

{\begin{lstlisting}
formulaUnits = 1 - 4
\end{lstlisting}}

In the new version of \texttt{XtalOpt}, introducing the chemical system and 
(optionally) desired formula units differs in the following ways from the previous versions:
(i) the ``empiricalFormula" and ``formulaUnits" pair of flags is removed, and
(ii) the chemical system is now entirely specified by the entries of a single input flag
``chemicalFormulas".

Generally, an input entry for ``chemicalFormulas":
\begin{enumerate}
\item Is interpreted as the ``explicit" formula of the simulation unit
cell (instead of the reduced empirical formula), and
\item Must include the ``full chemical formula"
(i.e., a combination of element symbols and corresponding
atom counts, e.g., ``Ti1O4" instead of ``TiO4").
\end{enumerate}

That is to say, the input:

{\begin{lstlisting}
chemicalFormulas = Ti4O8
\end{lstlisting}}

instructs \texttt{XtalOpt} to search in the Ti-O system with ``1:2" composition, while
all generated unit cells have four Ti atoms and eight O atoms.

Various ``formula units" of a chemical system can be introduced using a
comma-delimited list of full chemical formulas, e.g.,

{\begin{lstlisting}
chemicalFormulas = Ti1O2, Ti2O4, Ti3O6, Ti4O8, Ti8O16
\end{lstlisting}}

Besides explicitly listing various formula units in the chemical formulas input,
they can also be combined into a hyphen-separated list of formulas in the form of
a ``\textit{formula1 - formula2}" entry. In such an entry, both ends
must be full chemical formulas of the same composition, while ``formula2"
is a proper supercell of ``formula1". That is to say, the number of atoms of each element type
in ``formula2" must be a fixed integer multiple of those in ``formula1"
(e.g., ``A3-A7" does not work, while ``A3-A6" is accepted).

As an example, the above input can be specified using a hyphen-separated
entry as follows:

{\begin{lstlisting}
chemicalFormulas = Ti1O2 - Ti4O8, Ti8O16
\end{lstlisting}}

The above input corresponds to specifying the pair of input flags in the
previous versions of the code as:
``empiricalFormula = Ti1O2" and ``formulaUnits = 1-4, 8".

The ``chemicalFormulas" parameter not only determines the initial cell(s) to be
created in the first generation, it also has an important role in specifying what
type of the evolutionary search should be conducted, as detailed in the next section.

\section{Various types of evolutionary searches}

Generally, the overall search workflow of the new version of \texttt{XtalOpt} is
similar to the previous versions: an initial set of structures is
generated from the compositions and cell sizes specified in the input,
and the genetic operations are applied to the selected parents from the
pool to produce offspring.
However, now, the details of how the genetic operations make new offspring and the children that can be created are different. 
That is to say, depending on the search type, the
size and composition of the offspring produced by genetic
operations can be different than that of their parent structure(s) or
the input chemical formula list. The search type is determined by the
details of the input formulas list and a set of new flags, as described in the next subsections.

\subsection{Fixed-composition search}

In the simplest case, if all the input formulas are from the same
composition, the search is a ``fixed-composition" (FC) evolutionary search,
where the offspring have their composition match that of
their parent structure(s) (i.e., initial list). This is essentially the
traditional search performed by previous versions of \texttt{XtalOpt}.

\subsection{Multi-composition search}

In the new version of \texttt{XtalOpt}, on the other hand, the initial list
of chemical formulas can also include a combination of different
compositions, e.g.,

{\begin{lstlisting}
chemicalFormulas = Ti2O3, Ti1O2 - Ti4O8, Ti5O3, Ti1O1 - Ti5O5
\end{lstlisting}}

This input instructs \texttt{XtalOpt} to perform a ``multi-composition" (MC) search.
This type of search is similar to the FC search in that the 
initial set of structures is generated as usual, and subsequent individuals are forced to have a composition matching ``one of the compositions specified in the input list".
Since the initial list now covers
various compositions, however, the parent structure(s) can be chosen
from different compositions. This extends ``desired" motifs found in one
composition to other ones, potentially speeding up the search in finding
the best candidate structures for various compositions.

\subsection{Variable-composition search}

Regardless of whether the input formulas include one or more
compositions, if the input flag

{\begin{lstlisting}
vcSearch = true
\end{lstlisting}}

is specified, after the first generation is produced (as usual and
according to the input formulas list), the genetic operation crossover is now
allowed to produce cells with new compositions that do not
necessarily match their parent structures' composition,
or even those compositions listed among the initial chemical formulas.
This so-called ``variable-composition'' (VC) search is the most general type of evolutionary
search conducted by \texttt{XtalOpt}. In it the entire chemical system of the
specified elements is explored to find (meta)stable phases of various compositions.

\section{Minimum and maximum number of atoms}\label{limit}

Version 14 of \texttt{XtalOpt} can generate structures with new compositions
(hence, various numbers of atoms in the cell) and random supercells (discussed
below). To control the computational
cost of the search an input parameter, ``maxAtoms", is introduced
to set the maximum number of atoms in the generated unit cells during the search.
A new structure in the VC search or a random supercell will be
generated only if the total number of atoms in the final cell complies with
this set limit.

The default value for the maximum number of atoms is 20. However, and regardless
of being specified in the input or having its default value, if any entry in the input
chemical formulas list has a larger total number of atoms, \texttt{XtalOpt} will
re-adjust this parameter automatically to match the largest cell size in
the input.

In addition to the maximum atom counts setting, the user can
set a lower limit for the total atom counts in generated unit cells through the ``minAtoms" flag.
This parameter, which is relevant to a VC search, has the default value of 1.
If it is explicitly initialized to a value larger than the
smallest atom counts in the input chemical formulas, \texttt{XtalOpt} produces
a warning message and sets this parameter to the lowest possible value,
consistent with the input formulas.

\section{Genetic operations}

\texttt{XtalOpt} uses various genetic operations to produce new structures from
selected parent structure(s). In all of the previous versions, the defined operations
included crossover, stripple, and permustrain~\cite{Zurek:2011a}. Each operation would be
chosen based on their user-defined ``percent chance". The specified
chances for all operations, each an integer in the {[}0,100{]} range,
was required to sum to 100.

In \texttt{XtalOpt} version 14, the crossover operation is extended by allowing
for ``multi-cut" merging of the parent cells, and two new genetic operations
(applicable only to the VC search)
and a new random mutation (relevant to all types of
evolutionary searches), are introduced.
Moreover, the input entries defining the probability of choosing the various genetic operations and
the interpretation of their values have been changed, as described below.

\subsection{Multi-cut crossover} \label{sec:multi}

Prior to \texttt{XtalOpt} version 14, the crossover operation was performed by a single-point
``cut and splice" of the parent structures~\cite{Zurek:2011a}. The user now has the option to instruct the code to
cut the parent unit cells at multiple points to produce ``ribbons" by using the input flag:

\begin{lstlisting}
crossoverCuts = 3 # default is 1
\end{lstlisting}

Generally, if this flag is set to ``$n$", the code will cut each parent structure in ``$n$+1" ribbons,
and will select the atoms of the offspring structure from the alternating ribbons of the parent
structures. The width of these ribbons is initialized equally, and then distorted randomly. For the case
of the single-cut crossover, however, the setting for the minimum contribution of each parent is applied.
By default, \texttt{XtalOpt} will perform a single-cut crossover, unless the user instructs otherwise. This setting can be modified during the run.
It should be noted that specifying an ``even" number of cut points will result in an ``odd" number of ribbons, hence, an
uneven selection of atoms for the produced structure.

\subsection{New genetic operation: permutomic}

\begin{figure}[t]

\centering \includegraphics[width=.48\linewidth]{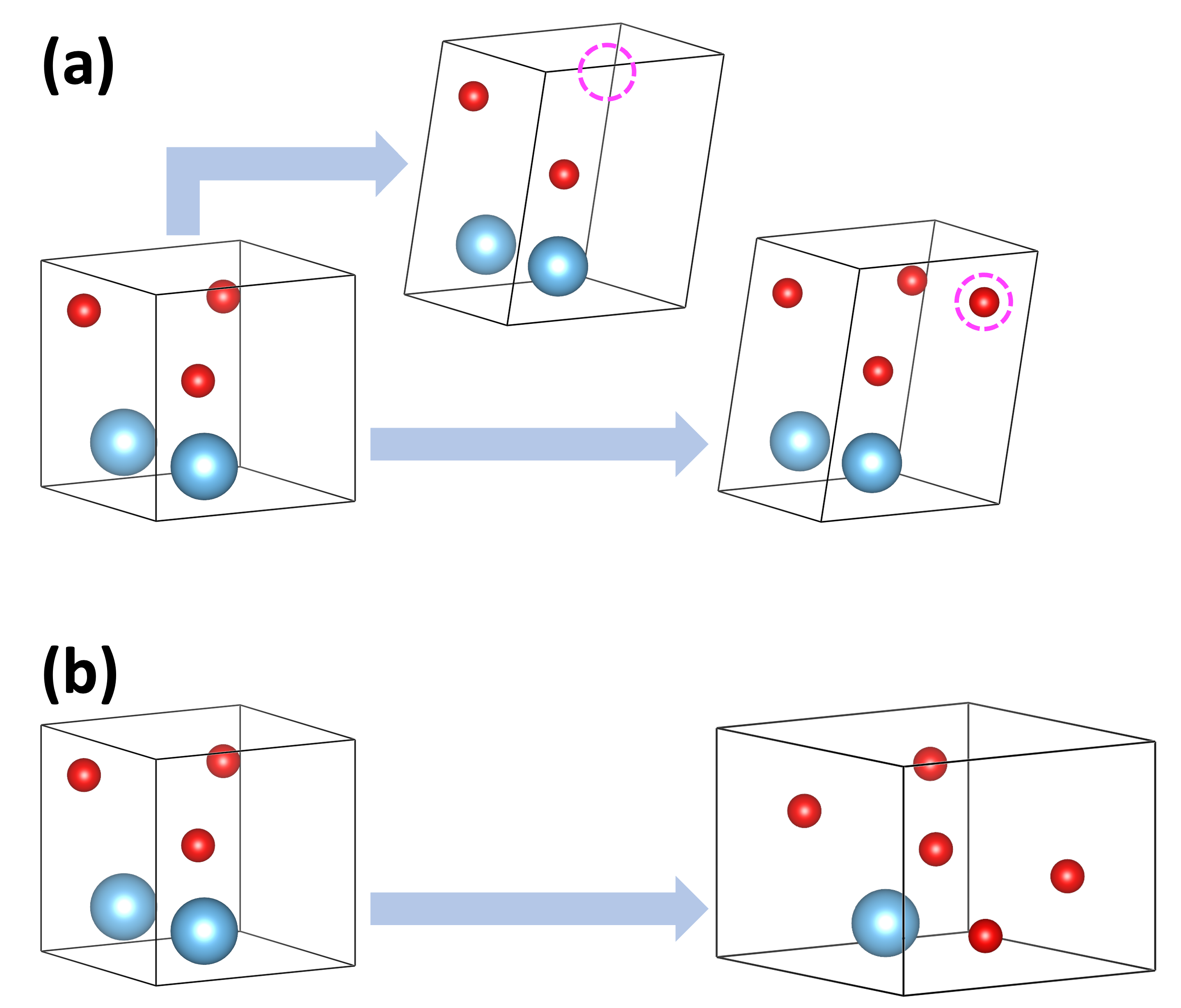}
\caption{Schematic illustration of the new genetic operations implemented in \texttt{XtalOpt} version 14, applicable
only to the VC search. (a) Permutomic produces an offspring
by randomly adding or removing an atom to the parent structure, followed by a small
random lattice distortion. (b) Permucomp is designed to diversify the
population through producing a new structure by distorting the parent lattice
and decorating it with a new random composition.}
\label{fig-permus}
\end{figure}

To ensure full sampling of the stoichiometry space, a new evolutionary operation, ``permutomic", has been introduced in the VC search.
This operation randomly adds (removes) an
atom to (from) a structure chosen from the parent pool, as schematically illustrated in Fig.\ \ref{fig-permus}(a).
For such a search, the user should provide a probability for performing this
operation in the input. This probability is adjustable during the run.

\subsection{New genetic operation: permucomp}

Another evolutionary operation introduced in the VC search to facilitate the exploration of various stoichiometries is ``permucomp". This operation, illustrated in Fig. \ref{fig-permus}(b), creates a new composition with a randomly chosen total number of atoms (up to ``maxAtoms" parameter).

This random mutation is primarily designed to diversify
the structure pool in long searches over multiple elements. Since it has a limited
chance of yielding an energetically favorable structure, ideally it should be
assigned a relatively small probability compared to other genetic operations.

\subsection{Chances of performing the genetic operations}

In \texttt{XtalOpt} version 14, instead of using a ``percent chance" for applying
the various genetic operations, the probability of choosing a particular operation
should be specified by a ``relative weight".
In the CLI input the relative weights can be specified using the
following set of flags (instead of the ``percentChances..." flag): 

{\begin{lstlisting}
weightCrossover
weightStripple
weightPermustrain
weightPermutomic
weightPermucomp
\end{lstlisting}}

The weight for each operation should be zero or any positive integer, and it can be modified during the run. It should be noted that no specific condition needs to be satisfied for the total sum of the weights (i.e.\ the sum does not need to equal 100). 

Assuming that the set of relative weights of the genetic operations are
given by \(\{ P_{i}\}\), \texttt{XtalOpt} determines the percent chance
of applying the \(j^\textrm{th}\) genetic operation, \(C_{j}\), as:

\[C_{j} = 100\  \times \ \frac{P_{j}}{\sum\limits_{i\textrm{ = all relevant operations}}^{} P_{i}}\]

The sum over relative weights includes genetic operations relevant to
the search: for a VC search it includes all five
genetic operations, while for a non-VC search the
weights of permutomic and permucomp are ignored and their chances are set to zero.

The default values for the relative operation weights, and the (approximate) corresponding
runtime percent chance for applying them for the various search types are
summarized in Table \ref{Tgens}.

\begin{table}[t!]
\small
\centering
\begin{tabular}{|l|l|ll|}
\hline
\multirow{2}{*}{Input flag} & \multirow{2}{*}{Default value} & \multicolumn{2}{l|}{Percent chance} \\ \cline{3-4}
                            &                                & \multicolumn{1}{l|}{VC}    & FC/MC  \\ \hline \hline
weightCrossover             & 35                             & \multicolumn{1}{l|}{33\%}  & 40\%   \\ \hline
weightStripple              & 25                             & \multicolumn{1}{l|}{24\%}  & 30\%   \\ \hline
weightPermustrain           & 25                             & \multicolumn{1}{l|}{24\%}  & 30\%   \\ \hline
weightPermutomic            & 15                             & \multicolumn{1}{l|}{14\%}  & 0\%    \\ \hline
weightPermucomp             & 5                              & \multicolumn{1}{l|}{5\%}   & 0\%    \\ \hline
\end{tabular}
\caption{Input flags to specify the relative weights of various genetic operations, with their
default values. The actual percentage chances of applying a genetic operation are listed in the last columns for the VC and FC/MC search types.}
\label{Tgens}
\end{table}

If all the relevant genetic operations for a specific type of search have zero
weight in the input file,
\texttt{XtalOpt} will use an equal probability for applying the genetic operations,
while a warning message is printed in the run output.

It should be noted that our benchmarking tests performed on the Ti-O system
show that the number of unique chemical compositions produced in a VC search increases
by 30\% or more with the use of permutomic and permucomp operations, with the default genetic
operation weights.

\subsection{Secondary mutation: random supercell expansion}

Although not a genetic operation \textit{per se}, the user can optionally define a
finite probability as a percent chance (in the {[}0, 100{]} range), by
which the structure generated from any of the existing genetic
operations will be expanded to a supercell. This probability is zero by
default, and can be set using the input flag:

{\begin{lstlisting}
randomSuperCell = 0
\end{lstlisting}}

\begin{figure}[t]
\centering
  \includegraphics[width=.48\linewidth]{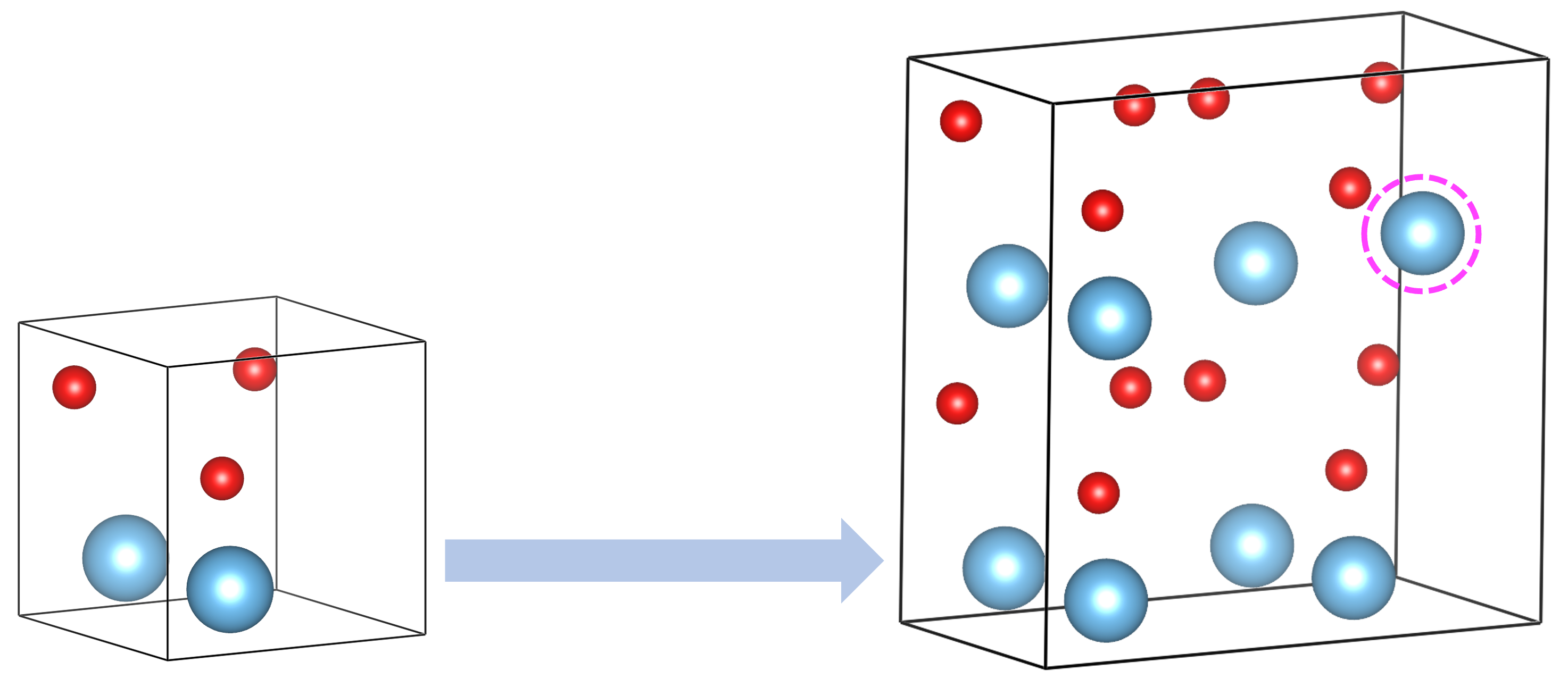}
\caption{The secondary mutation ``random supercell expansion" is applicable to the offspring produced
by any of the \texttt{XtalOpt} genetic operations, and produces a random supercell
in which an atom is randomly selected and displaced. The default probability of this mutation is 0\%.}
\label{fig-supercell}
\end{figure}

If this flag is specified to be a non-zero value, the expansion is performed by
randomly choosing three integers as multiples for the lattice vectors, such that the produced supercell has up
to ``maxAtoms" total atoms in the cell. Further, a randomly chosen atom in the supercell
is displaced randomly, subject to the minimum interatomic distances settings (Fig. \ref{fig-supercell}).
This operation can be useful in ground state searches where
charge density waves lower the energy of the system.

\section{Reference energies}

As of version 14, \texttt{XtalOpt} uses the distance above the convex hull as the
target value for global energy minimization, as first introduced by Zunger and co-workers~\cite{Trimarchi:2009a}. The calculation of the convex hull (performed using the \texttt{Qhull} library~\cite{Barber1996}),
requires defining the reference energies. By default, the code uses
``0.0" as the ``elemental reference energies". To obtain the correct convex hull for
MC and VC searches, however, the correct reference energies need to be specified.

The user can define reference
energies in the input through a comma-delimited list in which
each entry includes a full chemical formula followed by the corresponding
total energy (or enthalpy), e.g., for the Ti-O system:

{\begin{lstlisting}
referenceEnergies = O1 -2.53, Ti4 -17.65
\end{lstlisting}}

The energies introduced in the input must have the same unit as is used by the local optimizer (e.g.\ meV, eV or Hartree) for the chemical formula listed to produce the correct results.

Furthermore, the user can introduce reference energies 
not only of the elements but also for relevant multi-element (sub)systems if they
are needed for calculating the correct convex hull. An example of where this approach may be useful is if a particular binary or ternary is known to lie on the hull. For example, for the
O-Ti-N system, this can be a valid entry:

{\begin{lstlisting}
referenceEnergies = O2 -5.06, Ti2 -8.825, N1 -3, O1Ti1 -6.32, O2N5 -12.7, Ti1N2O1 -5.80
\end{lstlisting}}

It should be noted that if any reference energy is introduced, for
consistency, all elemental reference energies must be specified.
Otherwise \texttt{XtalOpt} issues an error message and quits.

\section{Pareto optimization and parent selection}

In release 13 of \texttt{XtalOpt}~\cite{SH12} a basic generalized
fitness function, calculated as the weighted sum of normalized values of objectives,
was introduced to perform a MOGO search for materials with desired functionalities. 
The effectiveness of this formalism has been demonstrated through
powder X-ray diffraction-assisted discovery of
meta-stable phases of Ti-O~\cite{SH14}, and the search for high-pressure
electride phases of Ca$_5$Pb$_3$ using electronic
charge density information~\cite{electrides}.
Further, it was previously shown that filtering the parent pool based upon
structural features that include the local crystalline order and symmetry is
an effective way to find metastable phases \cite{SH10}.

By default \texttt{XtalOpt} employs
this scalar fitness function for global optimization
(i.e., the ``Basic" optimization type) regardless of the number of objectives.
In the new version of \texttt{XtalOpt} Pareto optimization, which is in principle applicable to both SO and MO searches, has been implemented as an alternative (the ``Pareto" optimization type). 
The user can instruct the code to utilize
Pareto optimization through the input flag:

{\begin{lstlisting}
optimizationType = Pareto # default is Basic
\end{lstlisting}}

For Pareto optimization, \texttt{XtalOpt} uses the NSGA-II algorithm~\cite{Deb2002}
in which the Pareto front (rank) and crowding distances of the structures in the parent
pool are employed to select a new parent through a binary tournament selection.

The workflow of selecting a new parent structure in Pareto optimization
includes: (i) performing a standard non-dominated sorting to identify the
Pareto front (rank) of all candidate structures, (ii) calculating crowding distances
for structures in each front, (iii) randomly selecting a pair of structures from the pool,
and (iv) choosing the parent structure that has a better rank, or a larger crowding distance (if both structures belong to the
same Pareto front) from the selected pair. For structures with a similar rank and crowding distance,
the selection is made randomly.

Since \texttt{XtalOpt} is a population-based algorithm,
the parent pool includes all successfully optimized structures, hence,
elitism in the Pareto optimization is automatically maintained. Nevertheless,
the tournament selection is made within the entire population of
locally optimized structures after similar or duplicate (see Sec.\ \ref{sec:dup}) structures are identified, unlike the scalar fitness (basic) scheme where
a limited subset of the top candidate structures (i.e., equal to the size of the parent pool 
specified by the user) are considered for randomly selecting a parent
structure. The user, however, has the option to restrict the
tournament selection to the user-specified pool size by the flag: 

{\begin{lstlisting}
restrictedPool = true # default is false
\end{lstlisting}}

If the parent pool is restricted for tournament selection in Pareto optimization,
the code will sort the structures
in the entire pool according to their rank (and then according to the crowding
distances within each rank), and select the top candidates such that there will be
as many as the user-specified parent pool size from which
the tournament selection is performed.

Besides tournament selection, \texttt{XtalOpt} provides
another parent selection option within the Pareto scheme.  Here, the obtained rank and crowding distances of the structures
are used to calculate a scalar fitness, employed to determine the parents, similar to the case of basic optimization.  This can be invoked by using the flag:

{\begin{lstlisting}
tournamentSelection = false # default is true for Pareto optimization
\end{lstlisting}}

The Pareto-based fitness value relies on the calculated rank of structures,
$r$, which is a value in the $[0, N-1]$ range
, where $N$ is the total number of Pareto fronts and $r=0$ corresponds to
the global Pareto front.
Using the obtained ranks, solutions of each front are assigned a raw
fitness value of
\begin{equation}
f^{0}_{r} = \frac{N-r}{N}
\label{eq1}
\end{equation}

that corresponds to a fitness of $1$ for structures on the global
Pareto front and $\frac{1}{N}$ for those on the last Pareto front.
Next, the calculated crowding distances of structures
in each front are scaled to the $[0.1, 1]$ range,
where the non-zero lower limit is chosen
to eliminate the overlap between solutions of two successive fronts.

Finally, the Pareto-based fitness for
the $i^{th}$ structure in the $r^{th}$ Pareto front
with a scaled crowding distance of $d_{r,i}$
is calculated using the following formula:
\begin{equation}
f_{r,i} = f^{0}_{r} - \left( \frac{1 - d_{r,i}}{N}\right)
\equiv f^{0}_{r+1} + \frac{d_{r,i}}{N}
\label{eq2}
\end{equation}

The above procedure is designed to:
(i) ensure the rank-precedence by
introducing a gap of at least $\frac{0.1}{N}$ between the ``worst" solution
of rank $r$ and the ``best" solution of rank $r+1$,
(ii) diversify the selection pool
by prioritizing more unique candidate structures in each rank through applying
the crowding distance values, and
(iii) obtain a scalar fitness value in the range of $[0,1]$.

As outlined above, by default, \texttt{XtalOpt} applies crowding distances
in Pareto optimization in both tournament selection and Pareto-based fitness
calculation. This can be modified by specifying the input flag:

{\begin{lstlisting}
crowdingDistance = false # default is true
\end{lstlisting}}

which results in performing tournament selection only using the structural ranking,
or calculating Pareto-based scalar fitness with $d_{r,i}=1$ for all structures
(i.e., the raw fitness values obtained from Eq. \ref{eq1}).

It is important to note that the outcome of non-dominated sorting
is especially sensitive to the numerical precision of the objective values. Although
\texttt{XtalOpt} does not manipulate the calculated values of the objectives,
it provides the user with the option to apply a desired precision to them prior to the parent selection process. This may be done by specifying the number of decimal digits to be retained using the flag:

{\begin{lstlisting}
objectivePrecision = 6 # default is -1
\end{lstlisting}}

where the default value of ``$-1$" implies not rounding the objective values.

For an MO search performed using the previous version of \texttt{XtalOpt}~\cite{SH12},
specifying the objective weight and output
file name for the executable script was optional in the CLI mode. In the new
version of the code, these fields are all mandatory. Therefore, all four input entries
for an objective must be given in the below order:

{\begin{lstlisting}
objective = objective_type executable_script_path output_filename weight
\end{lstlisting}}

Furthermore, the weight for a ``Filtration" objective must be set to zero in the input.

It should be noted that:

\begin{itemize}
\item While the weights are not used for the Pareto optimization, they must always be
specified since the user has the option to change the optimization type during runtime
(Pareto to basic, and vice versa), and
\item Similar to the previous workflow, any minimizable/maximizable objective
with a weight of zero will be calculated but will not be considered in the
scalar fitness for the basic optimization scheme.
\end{itemize}

\section{Volume limits}

In the previous versions of \texttt{XtalOpt}, the volume of the unit cell could be
constrained by specifying either the absolute minimum and maximum limits
(in {\AA}$^{3}$ per formula unit), or the scaled volume limits.

In \texttt{XtalOpt} version 14, the absolute volume limits are read in 
``{\AA}$^{3}$ per atom" units and the corresponding flags are renamed to:

{\begin{lstlisting}
minVolume
maxVolume
\end{lstlisting}}

Also, the pair of flags for introducing the scaled volume limits are renamed to:

{\begin{lstlisting}
minVolumeScale
maxVolumeScale
\end{lstlisting}}

To be consistent with the minimum atomic radii settings, the scaled volume limits in the new version of \texttt{XtalOpt} are considered as factors of spheres of ``covalent" radii of atoms (whereas the previous version of the code used van der Waals radii). The scaled volume settings of the previous version,
as a result, need to be multiplied by $\sim$1.5 to obtain similar results in the new version.

Moreover, in the new version of \texttt{XtalOpt}, the user has the option
to define volume limits for the elements by providing a list of comma-delimited entries.
This setting will only take effect if the limits are defined for all elements in the chemical system.
Each entry includes a full chemical formula for
the elemental unit cell followed by the corresponding volume limits
(minimum and maximum, respectively) in units of {\AA}$^{3}$, e.g.,

{\begin{lstlisting}
elementalVolumes = O1 20 40 , Ti2 50 100
\end{lstlisting}}

Generally, the volume constraints in \texttt{XtalOpt} are applied in the
following order:

\begin{enumerate}
\item If elemental volumes are given (properly), they are used first,
\item If scaled factors are given (properly), they are used,
\item If explicit volume limits (per atom) are given, they are used,
\item If none of the above, the default absolute volume limits (1 and 100
{\AA}$^{3}$ per atom) are used.
\end{enumerate}

\section{Similarity check} \label{sec:dup}

Given that hundreds to thousands of structures are often optimized in a single evolutionary run, many of them end up being duplicates of one another. Removal of such duplicates from a search is of prime importance, otherwise the run can be biased by a particular structural motif, failing to explore the relevant areas of the potential energy landscape. However, duplicate identification is hampered by the fact that many structures are not fully optimized and because a single structure can be represented by different cell shapes and sizes. 

The first version of \texttt{XtalOpt} used an approximate scheme, identifying duplicates by comparing their spacegroups, volumes and enthalpies~\cite{Zurek:2011a} to within a user-specified tolerance. Because this method was prone to false positives, an exact comparison technique that searches for transformations mapping atoms from one structure to another was developed. The resulting \texttt{XtalComp} library~\cite{Zurek:2011i} was used to find duplicates since version 9~\cite{Zurek:2015h} of \texttt{XtalOpt}. Duplicates were marked as such in the output and removed from the run.  Though this exact matching scheme was key for finding certain inorganic compounds~\cite{Zurek:2014m}, it is dependent upon the choice of the user-specified tolerance and does not work well for molecular systems. For example, it would likely fail to flag structures that differ by a slight rotation of a molecule or a functional group as duplicates because the required tolerance would be too large. 

In these cases approximate methods that employ radial distribution functions (RDF) or atomic separation metrics to identify similar structures may be useful~\cite{Willighagen:de5009, Chisholm:wf5008, Abraham:2008}. Therefore, we have implemented a similarity check, which uses the RDF of species-resolved bonds to detect similarities between structures as first introduced in the \texttt{MAISE} code~\cite{SH07}, in \texttt{XtalOpt} version 14. Similar structures are labeled as such in the output. For instance, a label of ``\emph{Sim(2$\times$10)}" for the status of a structure in the \texttt{results.txt}
file means that this structure is similar to the structure ``2$\times$10",
within the given tolerances for the similarity check.

This functionality is controlled by a numerical value, in the {[}0,1{]} range,  set by the input flag:

{\begin{lstlisting}
rdfTolerance = 0.95
\end{lstlisting}}

If the user specifies a value greater than zero for the RDF tolerance,
\texttt{XtalOpt} will use the RDF similarity check instead of \texttt{XtalComp}. Specifically, it
calculates the scalar product of normalized RDF vectors of structures,
and those that have a value larger than the specified tolerance will be
marked as ``similar''.

By default, the RDF tolerance is set to zero; hence
\texttt{XtalComp} is employed for duplicate checking
to maintain consistency with previous versions of \texttt{XtalOpt}.
In general, however, the use of RDF check is recommended,
as it provides a quantitative measure for the similarity of structures that can
be visualized using common plotting software.

The details of the RDF vector calculations depend on a set of parameters, i.e.,

\begin{itemize}
\item Cutoff value for the included bond length (in {\AA}),
\item Spread of Gaussian functions used for smoothing the bond length distribution (in {\AA}),
\item Number of bins (over the bond length range of [0, cutoff]) for sampling the distribution.
\end{itemize}

In the CLI mode, this can be done through the following input flags (with their default values):

{\begin{lstlisting}
rdfCutoff = 6.0
rdfSigma = 0.008
rdfNumBins = 3000
\end{lstlisting}}

\section{Seed structures}

In previous versions of \texttt{XtalOpt} seed structures could be
introduced using a space-separated list. Now, for consistency, a comma-delimited list of entries should be used instead, e.g.,

{\begin{lstlisting}
seedStructures = /path1/POSCAR1,/path2/POSCAR2
\end{lstlisting}}

In previous versions the seeds were required to possess the same
composition as the reference chemical system of the search. However, in the new version of \texttt{XtalOpt}
a seed structure can be an ``off-composition" structure, i.e., a structure with a composition that is not found
in the input formulas list or is a sub-system of
the reference chemical system, such as binary or elemental structures in a ternary search.

The off-composition seeds will be read in as long as they do not include any element that
is not in the reference chemical system. They will be included in the parent pool
upon successful local optimization, and can participate in the
genetic operations if selected as a parent structure (depending on their fitness value). 
However, since the stripple and permustrain operations are
not designed to manipulate the parent structure's composition,
they will not be applied to an off-composition seed structure in the FC and MC searches.

One example where providing sub-system seeds could be useful is for obtaining energies of reference structures, to be subsequently used in the convex hull construction, that are consistent with the local optimization settings of the search. 
This option, however, should be utilized carefully since a failure of the local relaxation of the reference structure by the external code will result in an incorrect convex hull leading to errors in fitness determinations. The introduction of sub-systems may also be useful if they contain certain structural motifs that could be present in the full system.

The seed structures, in general, are not required to comply with the minimum and maximum number of atom settings.
Once a seed structure is read in with a total atom count that is not consistent with the existing settings, the
atom counts limits will be updated, while a warning message is produced in the output.

\section{Construction of molecular units}

For an MC or VC search, it is possible to use
molecular units for the construction of the initial random generation of structures, as first introduced in \texttt{XtalOpt} version 10~\cite{Zurek:2016k} for a FC search. However, now the number of atoms of each type that can participate in the construction of molecular units is limited to the smallest number of atoms of that type in the input list of chemical formulas, regardless of search type. 

For instance, given the following input:

{\begin{lstlisting}
chemicalFormulas = Ti2O4 - Ti4O8, Ti5O3
\end{lstlisting}}

up to 2 and 3 atoms of Ti and O, respectively, can be used
in the construction of molecular units. After initializing the molecular units in a unit cell,
the remaining atoms are added randomly to the cell to reach the desired composition.

\section{Interface to machine-learning interatomic potentials}

\texttt{XtalOpt} explicitly supports a number of optimizers whose output it can parse to find the structural coordinates and energies it requires. A simple way to employ an arbitrary optimizer (e.g., a machine learning interatomic potential) is to use simple scripting to convert its output to that of an already supported optimizer. 
For instance, if the user sets the optimizer type to \texttt{VASP} while employing an arbitrary
code to perform local optimizations, the job file for the \texttt{XtalOpt} run would include
the following steps: (i) converting the \texttt{VASP} structure file (\texttt{POSCAR}) to the appropriate format
for the optimizer employed, (ii) performing the local optimization, and (iii) extracting the results from
the optimizer and writing \texttt{VASP} format output (i.e., \texttt{OUTCAR} and \texttt{CONTCAR}) files.

Such a workflow would allow the user to benefit from the considerable speed-up that is offered by 
machine learning potentials in an \texttt{XtalOpt} run. This is especially helpful in
a VC search for multi-element systems where it might be necessary to explore the entire composition space. Such a run possibly involve thousands of local optimizations, which would be computationally prohibitive using first-principles approaches. Another possibility would be to first relax a structure locally with  a machine learning potential, followed by a DFT single point calculation to obtain a more accurate energy prediction.

An easy-to-use Python script, \texttt{vasp\_uip.py}, is included
in the new release of \texttt{XtalOpt} to facilitate such calculations.
This wrapper enables local optimizations with a series of the recently developed
universal interatomic potentials (UIPs), \texttt{MACE}~\cite{mace_arxiv},
\texttt{CHGNet}~\cite{Deng2023a}, 
\texttt{Orb}~\cite{rhodes2025orbv3}, and \texttt{MatterSim}~\cite{yang2024mattersim},
to perform local structure optimization
using the Atomic Simulation Environment (ASE) library~\cite{HjorthLarsen2017a}.
The input/output format is that of the \texttt{VASP} code, i.e.,
the script reads in a \texttt{POSCAR} file and produces a \texttt{CONTCAR} and a minimal \texttt{OUTCAR} file,
which follows the \texttt{VASP} format for the outputted entries.

The local optimization specifications and other parameters of this script can be
adjusted through a series of command line options. A full list of available
options can be obtained by typing:

{\begin{lstlisting}
python3 vasp_uip.py -h
\end{lstlisting}}

A simple calculation of the energies, forces, and stresses can be performed with this script as:

{\begin{lstlisting}
python3 vasp_uip.py
\end{lstlisting}}

provided a \texttt{POSCAR} file is present in the working directory, and the required libraries
are accessible through the invoked python binary.

With the above-mentioned input and output file formats,
this script is designed to be used in the \texttt{XtalOpt} runs as a \texttt{VASP} optimizer,
allowing for the efficient ground state search for a desired chemical system.
It should be noted that the implemented potentials have been chosen as prototypes
of the increasingly popular UIPs, and extending the script to support other similar
platforms is straightforward.

\section{Miscellaneous}

\subsection{Input entry conventions}

In a CLI run, various search settings are provided by the user in the input file,
through entries of pre-defined flags. While most flags require a single value as
the input entry, sometimes an entry is expected to include multiple parameters
(e.g., the ``objective" flag). There are also input flags that can parse
multiple entries, while the entries might or might not involve multiple parameters
(e.g., the ``elementalVolumes" and ``seedStructures" flags).

In general, and for consistency, the input patterns for the CLI flags are
designed with the following conventions:

\begin{itemize}
\item When input involves multiple entries, they should be a list of
\emph{comma-delimited} entries such as:

{\begin{lstlisting}
seedStructures = structure1 , structure2
\end{lstlisting}}

\item If a flag requires a single entry, and that entry has multiple
parameters, the parameters should be a list of \emph{space-separated} values, e.g.,
the specification of an objective (which requires four parameters):

{\begin{lstlisting}
objective = fil /bin/script output.txt 0.0
\end{lstlisting}}

\item If a flag accepts multiple entries that are multi-parameter,
combining the above rules, the input should be a \emph{comma-delimited
list of space-separated parameters}, e.g., specifying the
volumes for multiple elements (with three parameters for each elemental entry):

{\begin{lstlisting}
elementalVolumes = Ti1 20 35, O1 15 25
\end{lstlisting}}
\end{itemize}

The GUI input fields for chemical formulas, reference energies,
and elemental volumes should be initialized with input strings adhering to
the above format.

\subsection{Output files}

In \texttt{XtalOpt} 14 a new file, \texttt{hull.txt}, is 
created in the local run directory. It includes the composition
(atom counts of various elements), total enthalpy, and the calculated
distance above the hull for successfully optimized structures. Also
included in this file is the Pareto front index, creation index, and
a unique tag (generation and id) of each structure.

Moreover, with the exception of the \texttt{results.txt} and \texttt{xtalopt.state} files,
the names of the other output files (produced in the local run directory)
are changed in the new version of the code, as listed in Table \ref{Tfiles}.

\begin{table}[h!]
\small
\centering
\begin{tabular}{|l|l|}
\hline
Old filename                & New filename             \\ \hline \hline
xtalopt-runtime-options.txt & cli-runtime-options.txt  \\ \hline
xtaloptSettings.log         & settings.log             \\ \hline
xtaloptDebug.log            & output.log               \\ \hline
\end{tabular}
\caption{Old and new names of output files in an \texttt{XtalOpt} run.}
\label{Tfiles}
\end{table}

\subsection{Verbose output}

Starting from version 13 of \texttt{XtalOpt}, if the code was compiled with the
\texttt{cmake} flag

{\begin{lstlisting}
-DXTALOPT_DEBUG=ON
\end{lstlisting}}

the output of the run would include extra information about the run, and
this output would be saved to disk as the \texttt{xtaloptDebug.log} file (for both
the CLI and GUI modes). In \texttt{XtalOpt} release 14, for simplicity and regardless of
the compilation flag, the standard output of a CLI run will include an extensive set of
additional information (e.g., details of calculating objectives and parent selection probabilities,
dot product of the RDF vectors, workflow of genetic operations, etc.) by setting the following input flag:

{\begin{lstlisting}
verboseOutput = true # default is false
\end{lstlisting}}

This parameter has a default value of ``false", and it can be modified during the run.
It should be noted that, especially for a long search involving large unit cells,
this option can produce an enormous output file size; hence, it is most appropriate to be used
for debugging purposes.

In the GUI mode, this option is available as a check box in the ``Progress" tab
(Fig. \ref{fig-gui}(d)).
However, as in the previous version of \texttt{XtalOpt}, the \texttt{output.log} file
(that may or may not include the debug-type additional information) will be produced
only if the code is compiled with the above \texttt{cmake} flag. As
a result, and depending on the compilation type, this option in the GUI can be either
selectable or deactivated.

\subsection{Convex hull snapshots}
In the CLI mode, if the following flag is set:

{\begin{lstlisting}
saveHullSnapshot = true # default is false
\end{lstlisting}}

after each successful local optimization a snapshot of the hull data
(i.e., a copy of the \texttt{hull.txt} file) will be saved in the local
working directory under the folder ``movie" as a file named
\texttt{YYMMDD\_HHmmSS\_LLL} where: YY (year), MM (month), DD (day),
HH (hour), mm (minute), SS (seconds), and LLL (milliseconds) represent
a unique and ordered identifier of when the file was written.


This option is specifically designed to facilitate monitoring the progress of the run
or to create convex hull images and movies, directly from the file data or using external tools
such as the \texttt{pycxl} code~\cite{pycxlWebsite}.

In the GUI mode, the same option is available as the ``Save hull snapshots"
check box under the ``Search Settings" tab.

\subsection{Computational scaling}
In order to measure the computational cost of various \texttt{XtalOpt} tasks in a typical search, we performed
a series of benchmarking test runs on the Ti-O system to generate 100, 400, and 1000 structures. These searches were
conducted over a variety of input parameters, e.g., VC and non-VC searches, basic and Pareto optimization, with and
without additional objectives, and using RDF and XtalComp similarity checks. The runs were performed on 24 cores with
20 parallel local optimizations, and the \texttt{vasp\_uip.py} script with MACE UIP was employed for structural relaxations.

The profiling results were then used to measure the CPU time usage of various \texttt{XtalOpt} functions and tasks, and
compare them against the overall \texttt{XtalOpt} CPU usage and the total CPU time of the run, i.e., the total time
consumed by the code, external objective calculations, and the local optimizations that correlates with the run wall time.

Based on the above analysis, on average, \texttt{XtalOpt} consumes 35\%, 17\%, and 10\% of the total CPU
time in generating 100, 400, and 1000 structures, respectively. This is an expected result, as the increasing
number of generated structures in a search requires more CPU time for local optimizations (and objective
calculations, if applicable) performed by the external codes. A breakdown of the CPU time usage for various
tasks, averaged over all searches of different settings but the same target structure count,
is listed in Table \ref{tcc-1}. The results show that the overall process of the parent selection is the most
demanding task in a search, consuming on average $\sim$7\% of the code's CPU usage (0.9\% of the total CPU time).

\begin{table*}[h!]
\centering
\begin{tabular}{|l|l|l|}
\hline
Task                                  & Share of XtalOpt CPU time & Share of total CPU time \\ \hline \hline
Input/Output                          & 1.86\%                    & 0.38\%                  \\ \hline
Similarity check                      & 2.53\%                    & 0.53\%                  \\ \hline
Fitness analysis and parent selection & 6.95\%                    & 0.90\%                  \\ \hline
Structure generation and analysis     & 3.98\%                    & 0.77\%                  \\ \hline
\end{tabular}
\caption{Relative CPU time usage of various \texttt{XtalOpt} tasks, obtained by averaging the results of 
multiple profiling runs with different input settings; compared to the code and total CPU time usages.}
\label{tcc-1}
\end{table*}

Further, the most computationally-demanding \texttt{XtalOpt} functions were identified for various search
settings. The average share from the code and total CPU times for each function was calculated
over all searches with the same total number of generated structures. According to the average CPU time usages,
the non-dominated sorting and comparison of RDF vectors were found to be the most computationally expensive functions
in a typical search.
Currently, and by default, a non-dominated sorting is conducted regardless of the global optimization type to
obtain and output the Pareto front indices for user's information. This operation, on average, consumes 6\% of the
overall \texttt{XtalOpt} CPU usage (0.8\% of the total CPU time) in a typical search, ranking the function as the
most computationally expensive task. Second in the CPU usage is, when employed, the comparison of RDF vectors with
using 3.7\% of the overall \texttt{XtalOpt} CPU time (0.7\% of the total CPU time).

It should be noted that the share of computational cost of the code and its functions from the total CPU usage
of the runs is measured for searches conducted using fast UIP local optimizations.
Therefore, the computational overhead of the \texttt{XtalOpt} tasks would be insignificant
when the structural relaxations are performed using \textit{ab initio} codes.

\subsection{Legacy AFLOW-hardness optimization}

The explicit support for AFLOW~\cite{Curtarolo2012} hardness as an objective is removed.
AFLOW hardness can still be used as an objective, in the same way as  any objective that should be maximized, using an interface executable script. An example of such a script was introduced in the previous release of \texttt{XtalOpt}~\cite{SH12}.

\subsection{Backwards compatibility}

Version 14 of \texttt{XtalOpt} does not
read the input file
(\texttt{xtalopt.in}) of previous versions due to various changes in
the required input flags, and does not resume a run that was performed
with an older version of the code. However, old versions of the code can be readily obtained from the \texttt{XtalOpt} website~\cite{XtalOptWebsite} and the CPC program library.

\section{Conclusions}

In the present work, we outlined the latest developments of the \texttt{XtalOpt} evolutionary algorithm. The new version of the code expands the multi-objective global optimization (MOGO) feature beyond the generalized scalar fitness function by including the Pareto optimization algorithm. Further, a variable-composition search capability is implemented in the code, which allows the user to choose from various search types, i.e., searching for the ground state of a compound with a fixed composition, searching for (meta)stable phases over a specific set of pre-determined compositions, or exploring the full composition space of the target crystalline chemical system for locating stable phases of variable compositions.

The new release of \texttt{XtalOpt} also includes easy-to-use interface scripts for using modern universal interatomic potentials for local optimization of structures. With the described developments, bug fixes, and improvements to the code's workflow, \texttt{XtalOpt} version 14 is designed to offer the possibility of an efficient high-throughput exploration of the composition space for a chemical compound to predict novel (meta)stable phases with desired properties.

\section*{Acknowledgement}
We acknowledge the U.S. National Science Foundation
(DMR-2136038 and DMR-2119065) for financial support, and the
Center for Computational Research at SUNY Buffalo for computational support\\ (http://hdl.handle.net/10477/79221).


\newpage

\begin{appendices}
\renewcommand{\thetable}{\Alph{section}\arabic{table}}

\section{New implementation in the GUI}

\begin{figure*}[h!]
\centering
\includegraphics[width=.98\linewidth]{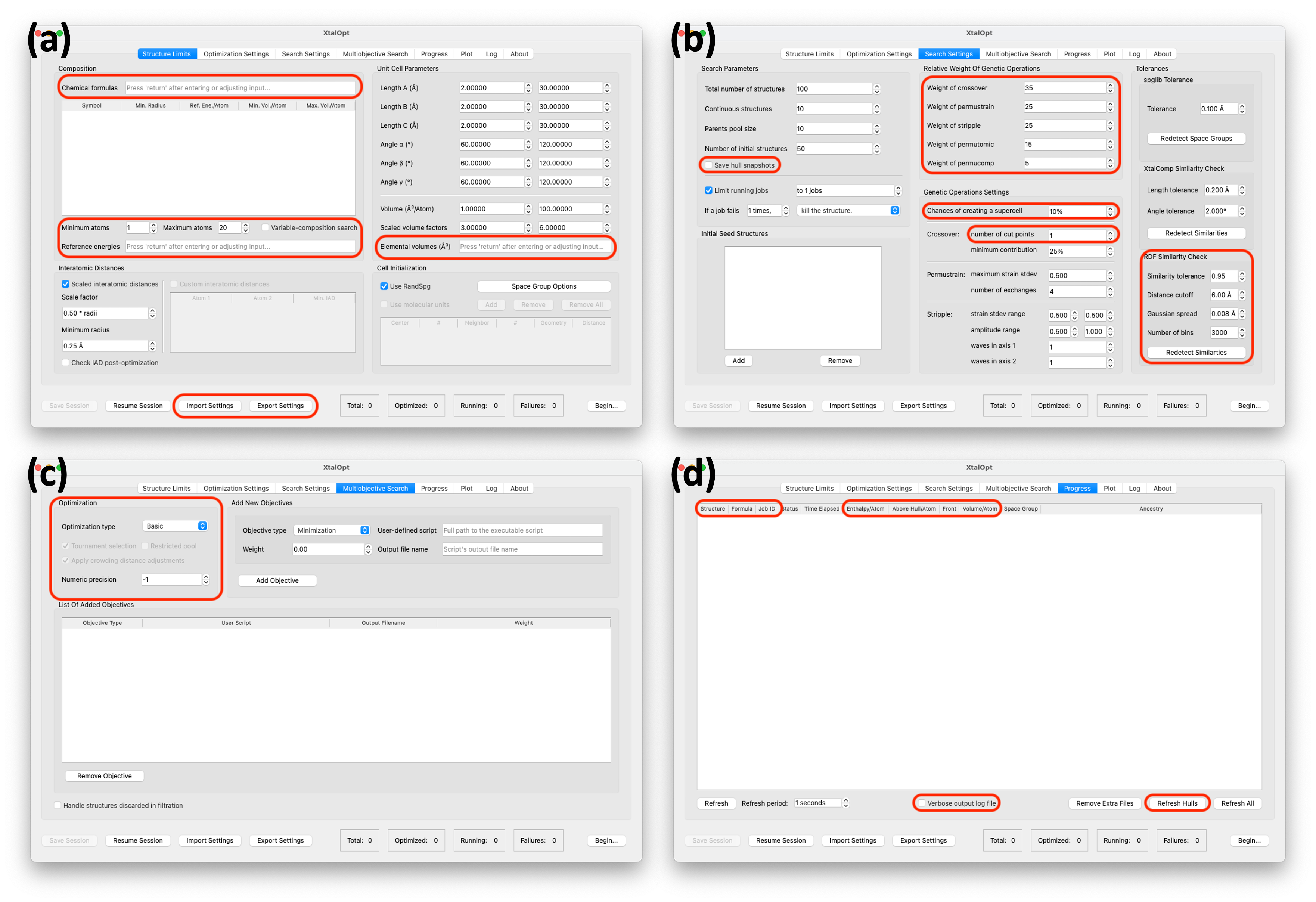}
\caption{The graphical user interface (GUI) of \texttt{XtalOpt} version 14 has been updated to include the new features:
(a) the ``Structure Limits" tab where the chemical system, search type, reference energies,
maximum number of atoms, various unit cell limits and cell initialization schemes
for the run can be set; (b) the ``Search Settings" tab entries can be used to specify the search duration
and run concurrency, action taken when a local optimization fails, initial seed structures,
genetic operation specifications, and thresholds for symmetry and similarity detections;
(c) the ``Multiobjective Search" tab parameters allow for adjusting the global
optimization type and its relevant settings, adding objectives for a
MO run, and specifying the code's actions when an external program fails to
calculate a filtration objective, and
(d) the ``Progress" tab maintains a live overview of the produced structures with
a summary of their characteristics, such as the chemical formula, enthalpy, distance above
the convex hull, Pareto front, symmetry, and ancestry information.}
\label{fig-gui}
\end{figure*}

The GUI of \texttt{XtalOpt} version 14 supports
all of the aforementioned developments. Generally, the format of the
relevant input entries (e.g., chemical formulas, reference energies,
elemental volumes) is the same as in the CLI mode. However, the GUI has been redesigned to
accommodate the changes made in this new implementation.
The ``Structure Limits" and ``Search Settings" tabs
host the majority of the new settings, while the ``Multiobjective Search"
tab has a new section (``Optimization") where the user can choose the
optimization type and adjust the relevant settings.
Also, the GUI now includes a new ``About" tab that offers some basic
information and web links where one can learn more about the code.
Fig. \ref{fig-gui} illustrates a few of the tabs in the new GUI,
with the changes relevant to the new implementation highlighted.

New to the GUI of \texttt{XtalOpt} is also the option to import run settings from a
CLI input file, and to export the GUI parameters to CLI settings file. These options, however,
are ``experimental" and provided for convenience. The user must verify that the imported parameters
in all GUI tabs (or entries in the exported CLI input file) match their expectation.

\newpage

\section{Summary of the new, modified, and obsolete CLI flags}

\begin{table*}[h]
\small
\centering
\begin{tabular}{|l|l|l|}
\hline
New flag             & Comments (default values indicated by ``*", if any)                     & Runtime adjustable\\ \hline \hline
chemicalFormulas     & Required input flag: cell formulas of initial generation                 & No\\ \hline
vcSearch             & Logical: perform variable-composition search (*false)                    & No\\ \hline
minAtoms             & Minimum unit cell size generated in the search (*1)                      & No\\ \hline
maxAtoms             & Maximum unit cell size generated in the search (*20)                     & No\\ \hline
referenceEnergies    & Cell formula and energy of convex hull references in local optimizer's units& No\\ \hline
weightPermutomic     & Relative weight of applying permutomic, applicable only to VC search (*15)& Yes\\ \hline
weightPermucomp      & Relative weight of applying permucomp, applicable only to VC search (*5)  & Yes\\ \hline
crossoverCuts        & Number of cut points for crossover operation (*1)                        & Yes\\ \hline
randomSuperCell      & Percent chance of performing random supercell expansion in [0, 100] (*0) & Yes\\ \hline
optimizationType     & Optimization scheme: Basic or Pareto (*Basic)                            & Yes\\ \hline
tournamentSelection  & Logical: use tournament selection in Pareto optimization (*true)         & Yes\\ \hline
restrictedPool       & Logical: restrict the tournament selection to top structures (*false)    & Yes\\ \hline
crowdingDistance     & Logical: apply crowding distances in Pareto optimization (*true)         & Yes\\ \hline
objectivePrecision   & Number of decimal digits in objective values for optimization (*-1)      & Yes\\ \hline
elementalVolumes     & Cell formula and volume limits for elements in {\AA}$^{3}$/cell units & Yes\\ \hline
rdfTolerance         & Threshold for RDF dot product to identify similarity in [0.0, 1.0] (*0.0)& Yes\\ \hline
rdfCutoff            & Maximum bond length considered for RDF vector in {\AA} (*6.0)            & No\\ \hline
rdfSigma             & Gaussian spread for RDF calculation in {\AA} (*0.008)                    & No\\ \hline
rdfNumBins           & Number of bins for RDF calculation (*3000)                               & No\\ \hline
saveHullSnapshots    & Logical: save snapshots of hull data (*false)                            & Yes\\ \hline
verboseOutput        & Logical: produce extra information in the run output (*false)            & Yes\\ \hline
user1                & Custom user-defined keyword                                              & No\\ \hline
user2                & Custom user-defined keyword                                              & No\\ \hline
user3                & Custom user-defined keyword                                              & No\\ \hline
user4                & Custom user-defined keyword                                              & No\\ \hline
\end{tabular}
\end{table*}

\begin{table*}[h!]
\small
\centering
\begin{tabular}{|l|l|l|l|}
\hline
Old flag                 & Renamed/changed flag  & Comments (default values indicated by ``*", if any)  & Runtime adjustable\\ \hline \hline
popSize                  & parentsPoolSize       &                                         & Yes\\ \hline
volumeMin                & minVolume             & Value should be in {\AA}$^{3}$ per atom & Yes\\ \hline
volumeMax                & maxVolume             & Value should be in {\AA}$^{3}$ per atom & Yes\\ \hline
volumeScaleMin           & minVolumeScale        & Factor of ``covalent" sphere            & Yes\\ \hline
volumeScaleMax           & maxVolumeScale        & Factor of ``covalent" sphere            & Yes\\ \hline
percentChanceCrossover   & weightCrossover       & Relative weight of applying operation (*35)  & Yes\\ \hline
percentChanceStripple    & weightStripple        & Relative weight of applying operation (*25)  & Yes\\ \hline
percentChancePermustrain & weightPermustrain     & Relative weight of applying operation (*25)  & Yes\\ \hline
objective                &                       & All entries are necessary               & No\\ \hline
seedStructures           &                       & A comma-delimited list                  & No\\ \hline
\end{tabular}
\end{table*}

\begin{table*}[h!]
\small
\centering
\begin{tabular}{|l|l|}
\hline
Obsolete flag(s)           & Comments                               \\ \hline \hline
empiricalFormula           & Replaced with ``chemicalFormulas" flag \\ \hline
formulaUnits               & Merged into ``chemicalFormulas" flag   \\ \hline
usingFormulaUnitCrossovers & It is performed by default             \\ \hline
usingOneGenePool           & It is performed by default             \\ \hline
mitosisA, mitosisB, mitosisC, printSubcell, & Mitosis feature and the corresponding flags are removed in      \\
mitosisDivisions, chanceOfFutureMitosis,    & \texttt{XtalOpt} version 14. See the ``random supercell expansion"  \\
usingMitoticGrowth, usingSubcellMitosis,    & (and ``randomSuperCell" flag) for a substitute functionality. \\ \hline
\end{tabular}
\end{table*}

\clearpage

\section{Example input files for an \texttt{XtalOpt} run}

\texttt{XtalOpt} input file for a VC search for the Ti-O system, using the MACE potential.

{\begin{lstlisting}[title={The content of the \texttt{xtalopt.in} file}]
softExit = true
chemicalFormulas = Ti7 O1, Ti1 O7
referenceEnergies = Ti3 -23.35901499 , O16 -78.97000885
vcSearch = true
maxAtoms = 40
numInitial = 50
parentsPoolSize = 20
limitRunningJobs = true
runningJobLimit = 2
continuousStructures = 15
maxNumStructures = 500
usingRandSpg = true
minVolumeScale = 3.0
maxVolumeScale = 6.0
weightCrossover   = 35
weightStripple    = 25
weightPermustrain = 25
weightPermutomic  = 15
weightPermucomp   = 5
randomSuperCell = 10
radiiScalingFactor = 0.5
aMin = 2.0
bMin = 2.0
cMin = 2.0
aMax = 30.0
bMax = 30.0
cMax = 30.0
alphaMin = 60.0
betaMin  = 60.0
gammaMin = 60.0
alphaMax = 120.0
betaMax  = 120.0
gammaMax = 120.0
templatesDirectory = ./templates
queueInterface = local
localWorkingDirectory = ./local
numOptimizationSteps = 1
optimizer = vasp
exeLocation = /Users/sam/TiO/vasp_uip.sh
incarTemplates = incar1
kpointsTemplates = kpoints1
potcarFile system = ./templates/potcar1
jobFailLimit = 1
jobFailAction = kill
rdfTolerance = 0.95
spglibTolerance = 0.01
\end{lstlisting}}

\newpage

Bash executable script to run the UIP interface script; to be used as a ``VASP"-like optimizer for \texttt{XtalOpt} run.
{\begin{lstlisting}[title={The content of \texttt{vasp\_uip.sh} script}]
#!/bin/bash

# Path to the UIP interface script
exe=/usr/local/bin/vasp_uip.py

# Load python environment
source /Users/sam/penv/bin/activate

# Relaxation task
python $exe -n 75 -o
\end{lstlisting}}

With the above files, and with a \texttt{templates/} folder that contains (empty) files \texttt{incar1}, \texttt{kpoints1},
and \texttt{potcar1}, one can run \texttt{XtalOpt} as:

{\begin{lstlisting}
/path/to/xtalopt --cli
\end{lstlisting}}

After the run is finished, the \texttt{pycxl} script can be used to plot the convex hull of the run, e.g.:

{\begin{lstlisting}
python /path/to/pycxl.py -i local/hull.txt -y -4 1 -c 1
\end{lstlisting}}

\begin{figure*}[h!]
\centering
\includegraphics[width=.50\linewidth]{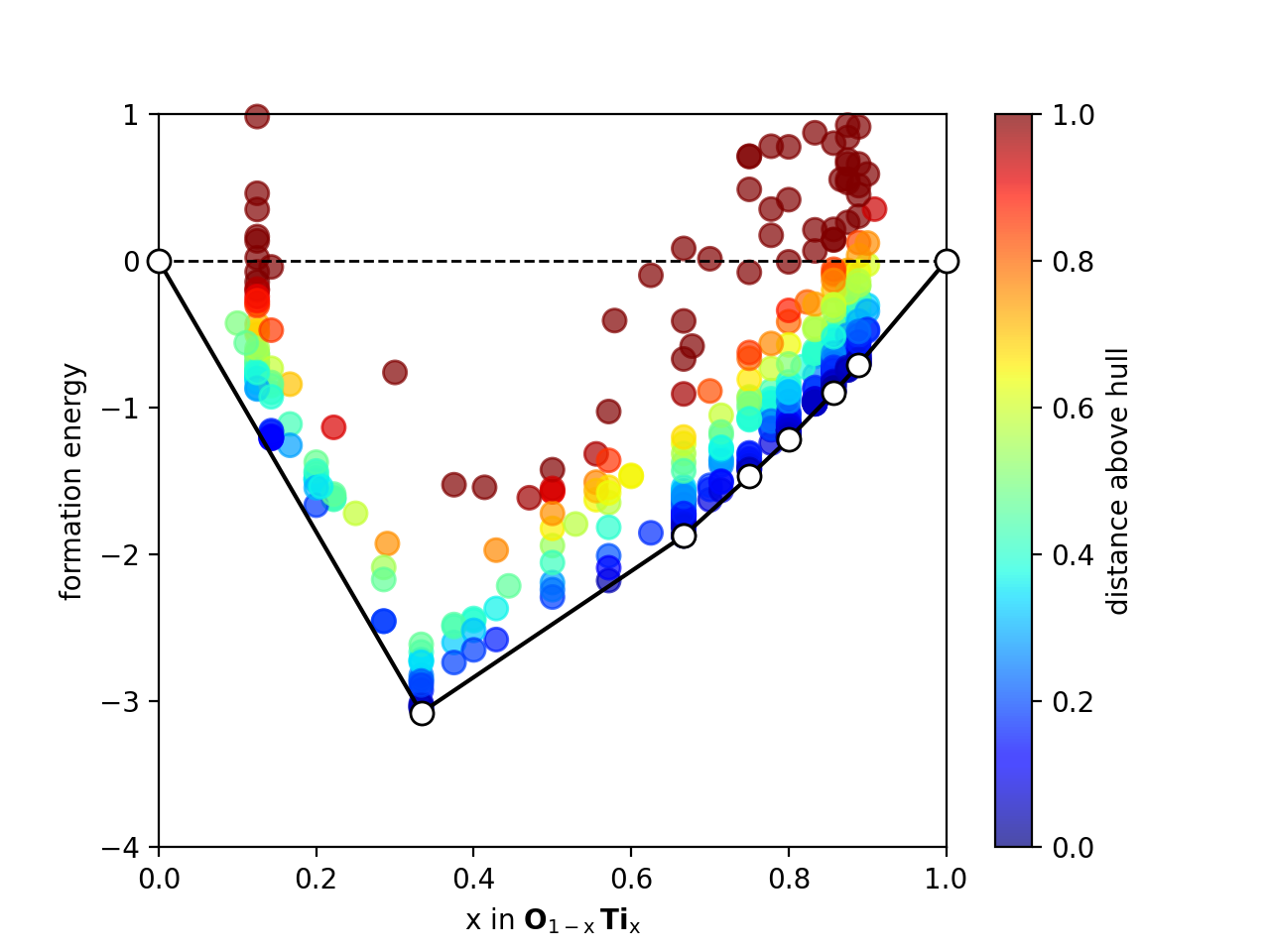}
\caption{An example of the convex hull produced by the \texttt{pycxl} code, from the output of the \texttt{XtalOpt} VC run
for the Ti-O system.}
\label{fig-chull}
\end{figure*}

This produces the \texttt{out\_distances.pdf} file, for which an example output is illustrated in Figure \ref{fig-chull}.

\end{appendices}

\clearpage



\bibliographystyle{elsarticle-num} 
\bibliography{library.bib}








\end{document}